\documentstyle[epsfig]{aipproc}
\begin{document}
%
%
%
%

\catcode`@=11
\def\chkspace{%
  \relax   
  \begingroup\ifhmode\aftergroup\dochksp@ce\fi\endgroup}
\def\dochksp@ce{%
  \unskip              
  \futurelet\chkspct@k\d@chkspc  
}
\def\d@chkspc{%
  \let\nxtsp@ce=\relax
  \ifx\chkspct@k.\else     
    \ifx\chkspct@k,\else
      \ifx\chkspct@k;\else
        \ifx\chkspct@k!\else
          \ifx\chkspct@k?\else
            \ifx\chkspct@k:\else
              \ifx\chkspct@k)\else
              \ifx\chkspct@k(\else
                \ifx\chkspct@k]\else
                  \ifx\chkspct@k-\else
                    \ifx\chkspct@k\egroup\else  
                      \let\nxtsp@ce=\put@space  
                    \fi
                  \fi
                \fi
              \fi
              \fi
            \fi
          \fi
        \fi
      \fi
    \fi
  \fi
  \nxtsp@ce
}
\def\put@space{$\;$}
\catcode`@=12

\def\Pade{Pad$\acute{\rm e}$\chkspace}

\def\ra{\relax\ifmmode \rightarrow\else{{$\rightarrow$}}\fi\chkspace}
\def\etal{{\it et al.}\chkspace}
\def\viz{{\it viz}\chkspace}
\def\adhoc{{\it ad hoc}\chkspace}
\def\ie{{\it i.e.}\chkspace}
\def\ap{{\it a priori}\chkspace}
\def\eg{{\it eg.}\chkspace}
\def\etc{{\it etc.}\chkspace}
\def\ala{{$\grave{a}\; la$}\chkspace}
\def\ibid{{\it ibid}\chkspace}
\def\defacto{{\it de facto}\chkspace}
\def\perse{{\it per se}\chkspace}
\def\apriori{{\it a priori}\chkspace}
\def\apost{{\it a posteriori}\chkspace}
\def\bonafide{{\it bonafide}\chkspace}

\def\ggx{{$\gamma\gamma$}\chkspace}
\def\sigg{{$\sigma_{\gamma\gamma}^{tot}$}\chkspace}
\def\fg{{$F^{\gamma}_2$}\chkspace}
\def\xg{{$x_{\gamma}$}\chkspace}
\def\syy{{$\sqrt{s}_{\gamma\gamma}$}\chkspace}
\def\gstar{{$\gamma^*\gamma^*$}\chkspace}
\def\be{\begin{equation}}
\def\ee{\end{equation}}
                         \def\bearr{\begin{eqnarray}}
                         \def\eearr{\end{eqnarray}}
\def\benum{\begin{enumerate}}
\def\eenum{\end{enumerate}}
\def\bitem{\begin{itemize}}
\def\eitem{\end{itemize}}

\def\ep{{e$^+$e$^-$}\chkspace}
\def\epa{{e$^+$e$^-$ annihilation}\chkspace}
\def\mup{{$\mu^+\mu^-$}\chkspace}
\def\taup{{$\tau^+\tau^-$}\chkspace}
\def\WW{{W$^+$W$^-$}\chkspace}
\def\qu{\quad}
\def\quu{\quad\quad}
\def\quuuu{\quad\quad\quad\quad}

\def\gluino{\relax\ifmmode \tilde{g} \else $\tilde{g}$ \fi\chkspace}

\def\qq{\relax\ifmmode q\overline{q}
\else $q\overline{q}$ \fi\chkspace}
\def\ff{\relax\ifmmode f\overline{f}
\else $f\overline{f}$ \fi\chkspace}
\def\gg{$\tilde{\rm g}\tilde{\rm g}$\chkspace}
\def\qbar{$\overline{\rm q}$\chkspace}
\def\bbar{$\overline{\rm b}$\chkspace}
\def\tbar{$\overline{\rm t}$\chkspace}
\def\QQ{Q$\overline{\rm Q}$\chkspace}
\def\pp{p$\overline{\rm p}$\chkspace}
\def\pbar{$\overline{\rm p}$\chkspace}

\def\bb{\relax\ifmmode b\bar{b}
       \else $b\bar{b}$ \fi\chkspace}
\def\ccrm{\relax\ifmmode {\rm c}\bar{\rm c}
       \else ${\rm c}\bar{\rm c}$ \fi\chkspace}
\def\cc{$c\bar{c}$ \chkspace}
\def\tt{\relax\ifmmode {\rm t}\bar{\rm t}
       \else ${\rm t}\bar{\rm t}$ \fi\chkspace}
\def\ss{\relax\ifmmode {\rm s}\bar{\rm s}
       \else ${\rm s}\bar{\rm s}$ \fi\chkspace}
\def\uu{\relax\ifmmode {\rm u}\bar{\rm u}
       \else ${\rm u}\bar{\rm u}$ \fi\chkspace}
\def\dd{\relax\ifmmode {\rm d}\bar{\rm d}
       \else ${\rm d}\bar{\rm d}$ \fi\chkspace}
\def\sbar{{$\bar{s}$}\chkspace}
\def\uds{{$u\bar{u},\;d\bar{d},\;s\bar{s}$}\chkspace}
\def\udsc{{$u\bar{u},\;d\bar{d},\;s\bar{s},\;c\bar{c}$}\chkspace}
\def\qqg{\relax\ifmmode q\overline{q}g
\else $q\overline{q}g$ \fi\chkspace}
\def\bbg{\relax\ifmmode b\overline{b}g
\else $b\overline{b}g$ \fi\chkspace}
\def\ccg{$c\overline{c}g$\chkspace}
\def\ttg{\relax\ifmmode t\overline{t}g
\else $t\overline{t}g$ \fi\chkspace}
\def\QQg{{Q$\overline{\rm Q}$g}\chkspace}
\def\qqgg{{q$\overline{\rm q}$gg}\chkspace}
\def\qqqq{{q$\overline{\rm q}${q$\overline{\rm q}$}}\chkspace}
\def\QQQQ{{Q$\overline{\rm Q}${Q$\overline{\rm Q}$}}\chkspace}
\def\bbbb{{b$\overline{\rm b}${b$\overline{\rm b}$}}\chkspace}
\def\cccc{{c$\overline{\rm c}${c$\overline{\rm c}$}}\chkspace}

\def\afb{\relax\ifmmode A_{FB} \else
{{$A_{FB}$}}\fi\chkspace}
\def\afbb{\relax\ifmmode A_{FB}^b \else
{{$A_{FB}^b$}}\fi\chkspace}
\def\pafb{\relax\ifmmode \tilde{A}_{FB} \else
{{$\tilde{A}_{FB}$}}\fi\chkspace}
\def\pafbb{\relax\ifmmode \tilde{A}_{FB}^b \else
{{$\tilde{A}_{FB}^b$}}\fi\chkspace}

\def\pafbzo{\relax\ifmmode \tilde{A}_{FB}|_{O(0)} \else
{{$\tilde{A}_{FB}|_{O(0)}$}}\fi\chkspace}
\def\pafbfo{\relax\ifmmode \tilde{A}_{FB}|_{\oalp} \else
{{$\tilde{A}_{FB}|_{\oalp}$}}\fi\chkspace}
\def\pafbso{\relax\ifmmode \tilde{A}_{FB}|_{\oalpsq} \else
{{$\tilde{A}_{FB}|_{\oalpsq}$}}\fi\chkspace}
\def\pafbto{\relax\ifmmode \tilde{A}_{FB}|_{\oalpc} \else
{{$\tilde{A}_{FB}|_{\oalpc}$}}\fi\chkspace}

\def\pafbbzo{\relax\ifmmode \tilde{A}_{FB}^b|_{O(0)} \else
{{$\tilde{A}_{FB}^b|_{O(0)}$}}\fi\chkspace}
\def\pafbbfo{\relax\ifmmode \tilde{A}_{FB}^b|_{\oalp} \else
{{$\tilde{A}_{FB}^b|_{\oalp}$}}\fi\chkspace}
\def\pafbbso{\relax\ifmmode \tilde{A}_{FB}^b|_{\oalpsq} \else
{{$\tilde{A}_{FB}^b|_{\oalpsq}$}}\fi\chkspace}
\def\pafbbto{\relax\ifmmode \tilde{A}_{FB}^b|_{\oalpc} \else
{{$\tilde{A}_{FB}^b|_{\oalpc}$}}\fi\chkspace}

\def\afbo0{\tilde{A}_{FB}|_{O(0)}}
\def\afbo1{\tilde{A}_{FB}|_{\oalp}}
\def\afbo2{\tilde{A}_{FB}|_{\oalpsq}}
\def\afbo3{\tilde{A}_{FB}|_{\oalpc}}

\def\lam{\relax\ifmmode \Lambda_{\overline{MS}}
       \else {{$\Lambda_{\overline{MS}}$}}\fi\chkspace}
\def\lamuds{\relax\ifmmode \Lambda^{(3)}_{\overline{MS}}
       \else {{$\Lambda^{(3)}_{\overline{MS}}$}}\fi\chkspace}
\def\lamudsc{\relax\ifmmode \Lambda^{(4)}_{\overline{MS}}
       \else $\Lambda^{(4)}_{\overline{MS}}$\fi\chkspace}
\def\lamudscb{\relax\ifmmode \Lambda^{(5)}_{\overline{MS}}
       \else $\Lambda^{(5)}_{\overline{MS}}$\fi\chkspace}
\def\alpb{$\alpha_s^{b}$\chkspace}
\def\alpc{$\alpha_s^{c}$\chkspace}
\def\alpbc{$\alpha_s^{bc}$\chkspace}
\def\alpuds{$\alpha_s^{uds}$\chkspace}
\def\alpall{$\alpha_s^{all}$\chkspace}
\def\alpudsc{$\alpha_s^{udsc}$\chkspace}
\def\alp{\relax\ifmmode \alpha_s\else $\alpha_s$\fi\chkspace}
\def\alpbar{\relax\ifmmode \bar{\alpha_s}
       \else $\bar{\alpha_s}$\fi\chkspace}
\def\alpmz{\relax\ifmmode \alpha_s(M_Z)\else $\alpha_s(M_Z)$\fi\chkspace}
\def\alpmzsq{\relax\ifmmode \alpha_s(M_Z^2)
       \else $\alpha_s(M_Z^2)$\fi\chkspace}

\def\oalp{\relax\ifmmode O(\alpha_s)\else{{O($\alpha_s$)}}\fi\chkspace}
\def\oalpsq{\relax\ifmmode O(\alpha_s^2)
           \else{{O($\alpha_s^2$)}}\fi\chkspace}
\def\oalpc{\relax\ifmmode O(\alpha_s^3)
           \else{{O($\alpha_s^3$)}}\fi\chkspace}
\def\oalpf{\relax\ifmmode O(\alpha_s^4)
           \else{{O($\alpha_s^4$)}}\fi\chkspace}

\def\rb{\relax\ifmmode R_3^b/R_3^{all}
           \else{{$R_3^b/R_3^{all}$}}\fi\chkspace}
\def\rc{\relax\ifmmode R_3^c/R_3^{all}
           \else{{$R_3^c/R_3^{all}$}}\fi\chkspace}
\def\ruds{\relax\ifmmode R_3^{uds}/R_3^{all}
           \else{{$R_3^{uds}/R_3^{all}$}}\fi\chkspace}
\def\ri{\relax\ifmmode R_3^i/R_3^{all}
           \else{{$R_3^i/R_3^{all}$}}\fi\chkspace}
\def\rj{\relax\ifmmode R_3^j/R_3^{all}
           \else{{$R_3^j/R_3^{all}$}}\fi\chkspace}
\def\alpi{\relax\ifmmode \alpha^i_s/\alpha^{all}_s
           \else{{$\alpha^i_s/\alpha^{all}_s$}}\fi\chkspace}

\def\mbz{\relax\ifmmode m_b(M_Z)
           \else{{$m_b(M_Z)$}}\fi\chkspace}
\def\mbb{\relax\ifmmode m_b(M_b)
           \else{{$m_b(M_b)$}}\fi\chkspace}

\def\plb{Phys. Lett.\chkspace}
\def\npb{Nucl. Phys.\chkspace}
\def\rmp{Rev. Mod. Phys.\chkspace}
\def\prl{Phys. Rev. Lett.\chkspace}
\def\prd{Phys. Rev.\chkspace}
\def\zpc{Z. Phys.\chkspace}

\def\z0{{$Z^0$}\chkspace}
\def\z0{\relax\ifmmode Z^0 \else {$Z^0$} \fi\chkspace}
\def\Dst{\relax\ifmmode {\rm D}^* \else {D$^*$}\fi\chkspace}
\def\Dpl{\relax\ifmmode {\rm D}^+ \else {D$^+$}\fi\chkspace}
\def\D0{\relax\ifmmode {\rm D}^0 \else {D$^0$}\fi\chkspace}
\def\Kst{\relax\ifmmode {\rm K}^* \else {K$^*$}\fi\chkspace}
\def\K0{\relax\ifmmode {\rm K}^0_s \else {K$^0_s$}\fi\chkspace}
\def\Kpl{\relax\ifmmode {\rm K}^+ \else {K$^+$}\fi\chkspace}
\def\Kstz{\relax\ifmmode {\rm K}^{*0} \else {K$^{*0}$}\fi\chkspace}

\def\beq{\begin{equation}}
\def\eeq{\end{equation}}
\def\bea{\begin{eqnarray}}
\def\eea{\end{eqnarray}}

\title{Two Photon Physics at Future Linear Colliders} 

\author{A. De Roeck$^*$}
\address{$^*$ CERN 1211 Geneva 23 Switzerland \\ 
e-mail: deroeck@mail.cern.ch}
\maketitle

\begin{abstract}
Prospects for QCD studies in two-photon interactions at a future linear 
\ep and \ggx collider are discussed. 
\end{abstract}

\section{Introduction}

Traditionally \ep colliders provide a wealth of two-photon data.
The photons are produced via bremsstrahlung~\cite{wws} from the
electron and positron 
beam, which  leads to a soft energy spectrum for the photons.
Such processes will also occur at future high energy (0.5-1 TeV)
\ep colliders,
but due to the ``single time'' usage of the colliding beams
these will allow other operation modes, such as a photon collider mode.  
A photon collider~\cite{ginz,telnov}, where 
the  electron beams  of a  linear \ep collider are 
converted into  photon
beams
via Compton laser backscattering, offers an exciting possibility to study
two-photon interactions at the highest possible energies,
with high luminosity. A plethora of
QCD physics 
topics in two-photon interactions can be addressed 
with a linear \ep collider or photon collider.
In this mini review we discuss perspectives for measurements of the total 
cross-section, the photon structure function and the onset of large logarithms
in the QCD evolution.

\section{Total Cross Section}
\begin{figure}[htbp]
\begin{center}
\epsfig{figure=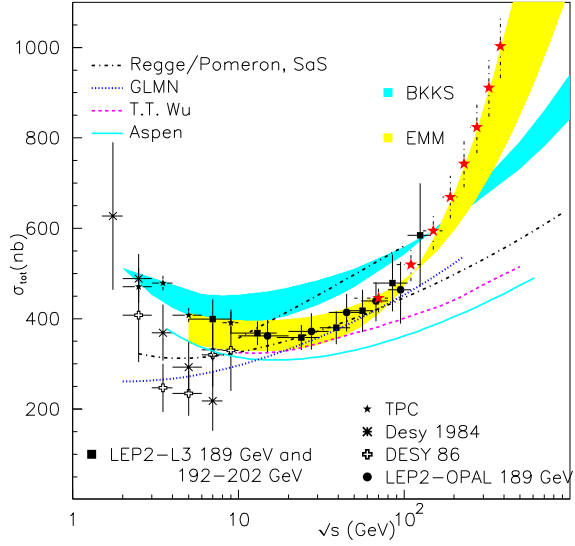,bbllx=0pt,bblly=150pt,bburx=540pt,bbury=580pt,width=8cm}
\caption{
The total \ggx cross-section as function of the \ggx collision energy, 
compared with model calculations:
BKKS band (upper and lower limit correspond to different   
photon densities);
SAS lines (Regge Pomeron exchange, upper and lower limits 
as given by SAS); 
Aspen (QCD inspired model, satisfying factorization);
EMM band (Eikonal Minijet Model for total and 
inelastic cross-section, with different photon densities and 
different minimum jet transverse momentum).
}
    \label{tdr-phys-qcd:fig_ggtot}
\end{center}
\end{figure}

 The  total \ggx cross-section
is not yet understood from first principles.
Fig.~\ref{tdr-phys-qcd:fig_ggtot} shows 
 the present photon-photon cross-sections data
in comparison with recent phenomenological 
models~\cite{pancheri}.
All models predict a rise of the cross-section with the collision 
energy, \syy, but the amount of the rise
differs and 
predictions for high photon-photon  energies show dramatic
differences. 
In {\it proton-like-models}
(dash-dotted~\cite{SAS,us1}, 
dashed~\cite{ttwu},dotted~\cite{GLMN} and 
solid~\cite{aspen} curves), the curvature 
follows  closely that of proton-proton cross-section, while in {\it QCD based} 
models (upper~\cite{BKKS} and lower~\cite{pancheri,us1} bands), the rise is 
obtained using the eikonalized PQCD jet cross-section. 

The figure demonstrates that large differences between the models 
become apparent in the energy  range  
of  a future 0.5-1 TeV \ep collider. 
A detailed comparison of the 
predictions~\cite{pancheri} reveals that in order to distinguish between 
all the models the cross-sections  need to be determined 
to a precision of better than 10\%. This is difficult to achieve 
in the \ep collider mode, since the variable \syy needs to be 
reconstructed from the visible hadronic final state in the detector. 
At the highest 
energies, the hadronic final state extends in pseudorapidity
$\eta = \ln\tan\theta/2$ in the region $-8 < \eta < 8$, while 
the detector covers roughly  the region $-3 < \eta <3$. 
However, for a photon collider the photon 
beam energy can be tuned with a spread of less than 10\%, such that
measurements
of \sigg can be made at a number of different energy values in the range
$50 <$ \syy $< 400$ GeV. The absolute precision with which
these cross-sections can be measured ranges from 5\% to 10\%,
where the largest contributions to the errors are due to 
the control of the diffractive component  of the cross-section,
Monte Carlo models used to correct for the event selections,
the absolute luminosity and knowledge on the 
shape of the luminosity spectrum.
It will be necessary to constrain the diffractive component
in high energy two-photon data. A technique
to measure diffractive
contributions separately, mirrored to 
the rapidity gap methods used at HERA,
 has been proposed in \cite{engel}.



\section{Photon Structure}
The nature of the photon is complex. A high energy photon can fluctuate
into   a fermion pair or even into a bound state, i.e. a vector meson with
the
same quantum numbers as the photon $J^{PC} = 1^{--}$.
These  quantum fluctuations 
lead to the
so-called hadronic structure of the 
photon.
In contrast to the structure function of the proton the structure 
function of the photon is predicted to rise linearly with the 
logarithm of the momentum transfer $Q^2$, and to increase with
increasing Bjorken-$x$~\cite{gg_zerwas}. The absolute magnitude 
of the photon structure function is asymptotically determined by 
the strong coupling constant~\cite{gg_witten}.

The classical way to study the structure of the photon is via
deep inelastic electron-photon scattering, i.e. two-photon interactions 
with one quasi-real (virtuality $Q^2 \sim 0$) and one virtual
($Q^2 >$ few GeV$^2$) photon.
The unpolarised $e\gamma$ DIS cross-section is
\begin{equation}
\label{avs1}
  \frac{d \sigma (e \gamma \rightarrow eX)}{dQ^2 dx}
  \: = \:\frac{2\pi \alpha^2}{Q^4 x} 
  \:\cdot 
  \Big[ \big\{ 1 + (1-y)^2 \big\} F_2^{\, \gamma}(x,Q^2) 
  - y^2 F_L^{\, \gamma}(x,Q^2) \Big] \: , 
\end{equation}
where $F_{2,L}^{\, \gamma}(x,Q^2)$ denote the structure functions of the
real photon.

 \begin{figure}[htbp]
\begin{center}
\epsfig{figure=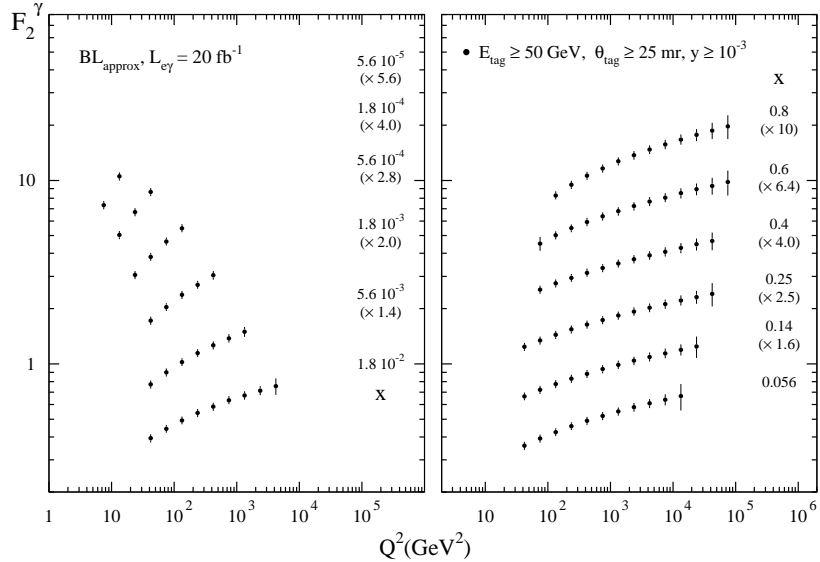,bbllx=80pt,bblly=30pt,bburx=530pt,bbury=720pt,angle=-90,width=11cm}
\caption{The kinematic coverage of the measurement of \fg for the 
backscattered $e\gamma$ mode at a 500 GeV linear collider.}
    \label{tdr-phys-qcd:fig_ggf2}
\end{center}
\end{figure}

To measure \fg it is important to detect (tag) the scattered 
electron which has emitted the virtual photon. Background studies 
suggest that these electrons can be detected down to 25 mrad and down
to 50 GeV.  
$e\gamma$ scattering at a photon collider resembles experimentally
$ep$ scattering at HERA, i.e. the energy of the probed quasi-real photon  is 
known (within the beam spread of 10\%) and the systematic error can be
controlled to about 5\%.
Fig.~\ref{tdr-phys-qcd:fig_ggf2}
shows the measurement potential for a photon collider~\cite{vogt}.
The measurements are shown with statistical and (5\%)
systematical error, for 20 fb$^{-1}$ photon collider luminosity, i.e.
about a year of data taking. Measurements can be made in the region 
$5.6 \cdot 10^{-5} < x < 0.56$, i.e. in a region similar to the HERA proton
structure function measurements, and $10 < Q^2 < 8\cdot 10^{4}$ GeV$^2$.
For the \ep collider mode the hadronic final state needs to be measured
accurately in order to reconstruct $x$. This will limit
  the lowest reachable $x$ value 
around $10^{-3}$. It will enable however measurements in the 
high $x$ ($0.1 < x < 0.8$) and high $Q^2$ ($ Q^2> 100$ GeV$^2$) 
range, for detailed \fg QCD evolution tests\cite{nisius}.


The $Q^2$ 
evolution of the structure function
at 
large $x$ and $Q^2$ has also been often advocated as a clean measurement
of $\alpha_s$. A 5\% change on $\alpha_s$ results however in a 
3\% change in \fg only, hence such a 
$\alpha_s$ determination  will require very precise \fg
measurements.

At high $Q^2$ values, apart from $\gamma$ exchange,
 also $Z$ and $W$
exchange will become important, 
the latter leading to charged current events~\cite{ridder}
which leads to spectacular signals due to the escaping
neutrino with high transverse momentum. By
measuring the electroweak neutral and charged current structure functions,
the up and down type quark content of the photon can be determined
separately.

\begin{figure}[htbp]
\begin{center}
\epsfig{figure=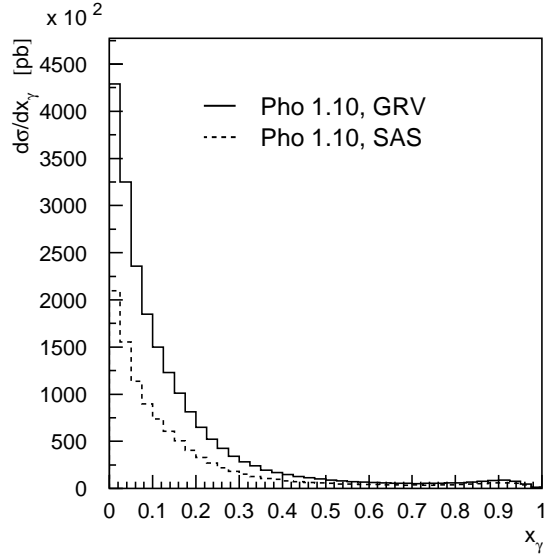,bbllx=30pt,bblly=0pt,bburx=410pt,bbury=380pt,width=8cm}
\caption{Jet cross-sections versus \xg for the 
backscattered $\gamma\gamma$ mode at a 500 GeV linear collider, 
for two assumptions of parton distributions of the photon.}
\label{tdr-phys-qcd:fig_ggjet}
\end{center}
\end{figure}

While $e\gamma$ scattering allows to measure the quark distributions
it only constrains the gluon distribution via the QCD evolution
of the structure functions. Direct information on the gluon in the 
photon can however be obtained from measurements of jet and
charm production~\cite{jankowski} in 
$\gamma\gamma$ collisions at an \ep or $\gamma\gamma$ collider. 
Fig.~\ref{tdr-phys-qcd:fig_ggjet} shows the Di-jet cross-section as 
function of \xg $= x^{\pm}_{\gamma}=\Sigma_{jets}(E\pm p_z) /
\Sigma_{hadrons}(E\pm p_z)$, with $p_z$ the longitudinal 
momentum of a particle. This variable is closely related to the 
true \xg at the parton level, and can be used to separate 
resolved (e.g. $x^{\pm}_{\gamma}< 0.8$) from direct 
(e.g. $x^{\pm}_{\gamma}> 0.8$) processes. 
The \xg distribution is shown
for two different assumptions of the parton distributions in dijet 
production. \xg values down to a few
times 10$^{-3}$ can be reached with 
charm and di-jet measurements~\cite{wengler}.

 \begin{figure}[htbp]
\begin{center}
\epsfig{figure=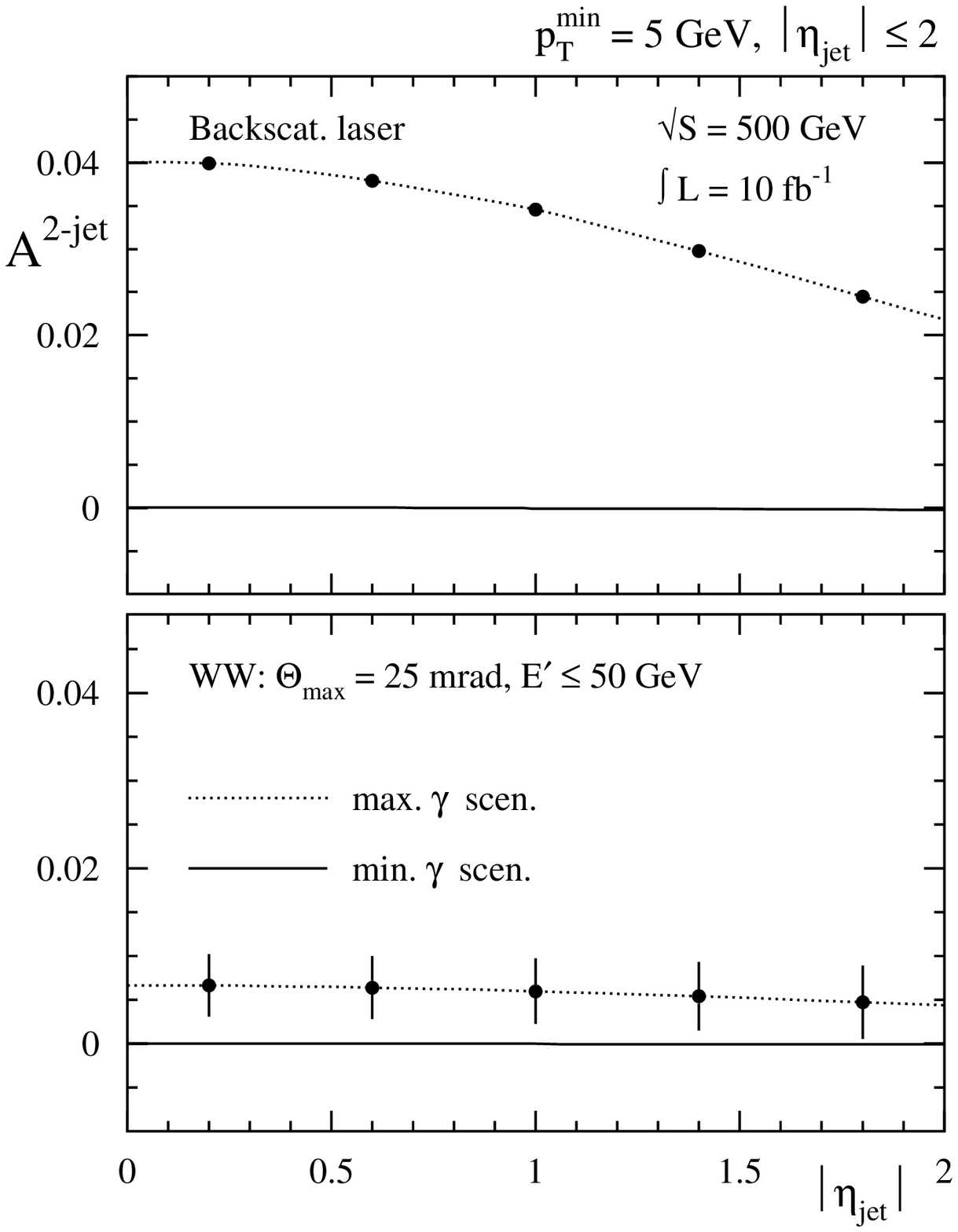,bbllx=0pt,bblly=0pt,bburx=430pt,bbury=580pt,width=6cm}
\epsfig{figure=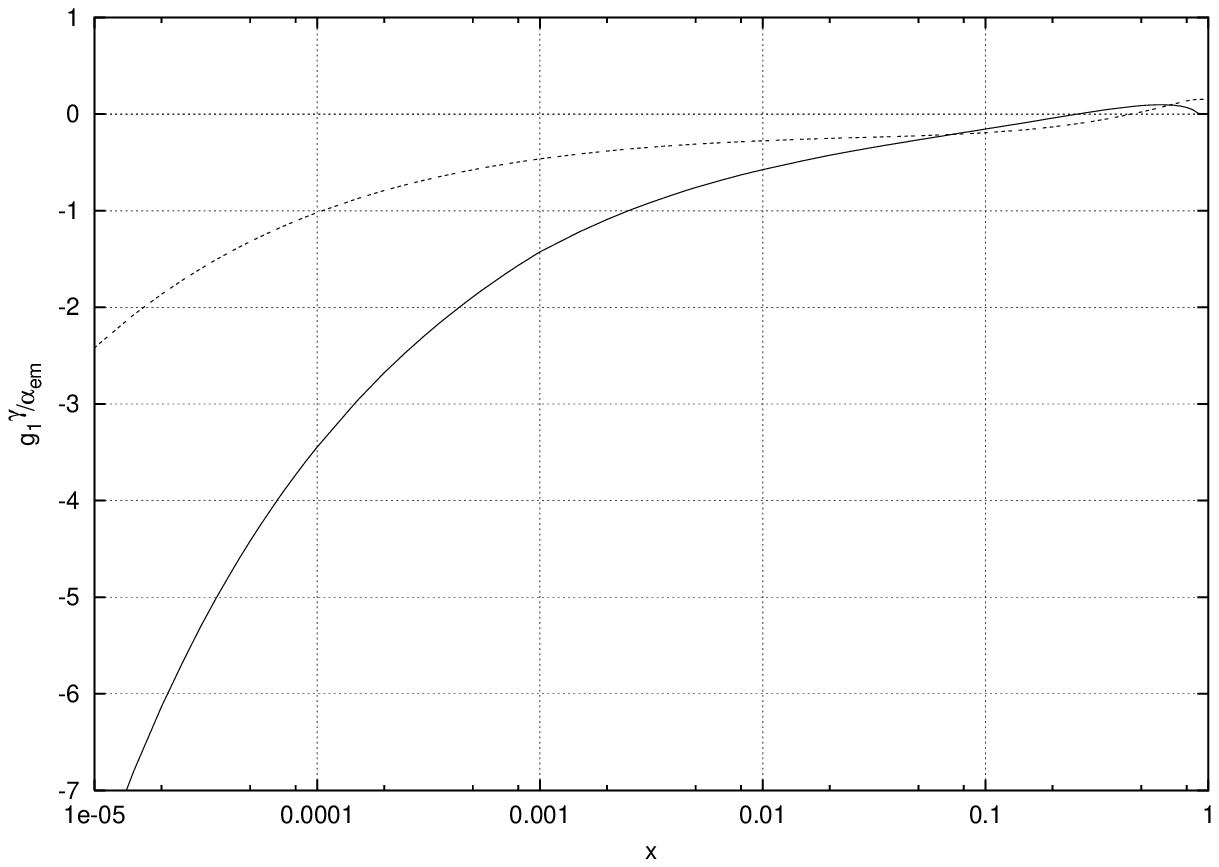,bbllx=40pt,bblly=0pt,bburx=400pt,bbury=300pt,width=6cm,clip=}
\caption{(left) di-jet spin asymmetry for events with $x^{\pm}_{\gamma} < 0.8$,
$p^{jet}_T = 5$ GeV and $|\eta_{jet}|< 2$ for $e\gamma$ (top) and
$\gamma\gamma$ (bottom) collisions. Predictions are shown for two different 
assumptions for the polarized parton distributions of the photon.
(right) Predictions for $g_1$ calculated without (dashed line) and 
with (full line) $\alpha_s^n\ln^{2n}1/x$ effects. }
\label{tdr-phys-qcd:fig_ggpol}
\end{center}
\end{figure}

A linear collider also provides circularly polarised 
photon beams, either from the polarised beams of the  \ep collider directly,
or via polarised laser beams scattered on the polarised \ep drive beam.
This offers a unique opportunity to study the 
polarised parton distributions of the photon, for which no 
experimental data are available to date. 


Information on the spin structure of the photon can be obtained from 
inclusive polarised deep inelastic $e\gamma$ measurements and from 
jet and charm measurements~\cite{stratmann,g1_kwiecinski}
in polarised \ggx scattering. 
An example of a jet measurement is presented in 
Fig.~\ref{tdr-phys-qcd:fig_ggpol} which shows the asymmetry measured 
for dijet events, for the \ep and photon collider modes 
separately. Two extreme
models are assumed for the polarised parton distributions in the 
photon.
Already with very modest luminosity significant measurements of the 
polarised parton distributions become accessible at a linear collider.
The extraction of the polarised structure
function $g_1 (x,Q^2) =\Sigma_q e^2_q( \Delta q^{\gamma}(x,Q^2)+ 
\Delta\overline{q}^{\gamma}(x,Q^2))$, with $\Delta q$ the polarised
parton densities, can however be best done at a $e\gamma$  collider.
Measurements of $g_1$, particularly at low $x$,
 are  extremely important for studies of the high energy QCD
limit, or BFKL regime~\cite{bfkl}. 
Indeed, the most singular terms of the effects of the  small $x$ resummation
on $g_1(x,Q^2)$
behave like $\alpha_s^n\ln^{2n}1/x$, compared to $\alpha_s^n\ln^{n}1/x$
in the unpolarised case of \fg. 
Thus large $\ln 1/x$ effects are expected to set in 
much more rapidly for polarised than for unpolarised structure measurements.
Fig.~\ref{tdr-phys-qcd:fig_ggpol} shows that for
leading order calculations, including kinematic constraints, 
the differences in predictions for $g_1$ 
with and without these large logarithms can be as large as a factor 
 3 to 4 for $x= 10^{-4}$ and could thus be easily measured with a few
years of data taking at a photon collider.

\section{Testing of BFKL Dynamics}
Apart from the inclusive polarised structure function 
measurements, discussed in the 
previous section, several dedicated measurements exist for 
detecting and studying the large $\ln1/x$ logarithm 
resumation effects in QCD, also called BFKL dynamics.

\begin{figure}[htbp]
\begin{center}
\epsfig{figure=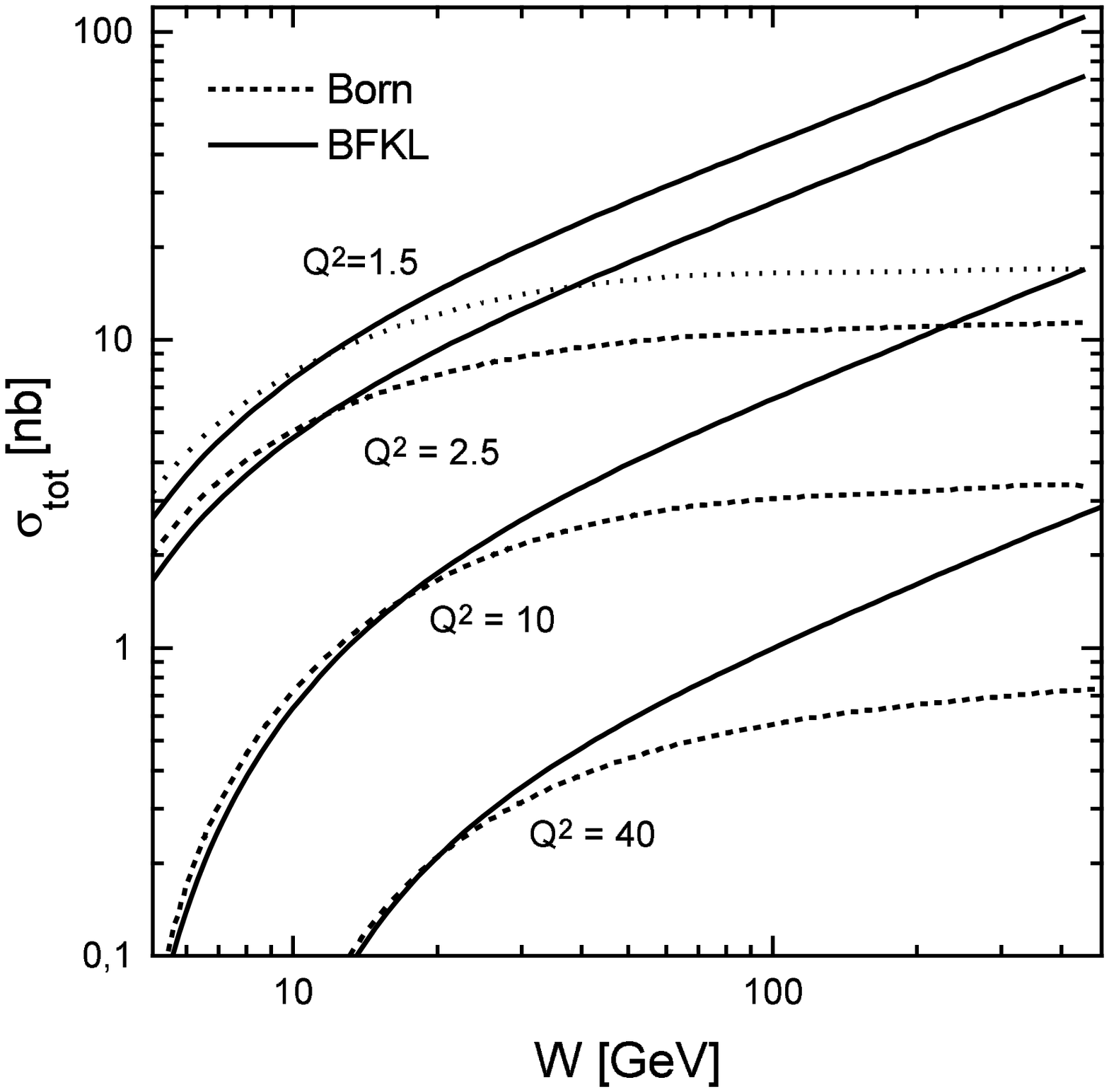,bbllx=0pt,bblly=220pt,bburx=600pt,bbury=740pt,width=9cm}
\caption{Prediction of the $\sigma_{\gamma*\gamma*}(Q^2,Q^2,W^2)$ 
cross section 
(solid line)
and two gluon exchange cross section (dotted line) as function 
of $W^2$ for different $Q^2$ values.
}
\label{tdr-phys-qcd:fig_gsgs}
\end{center}
\end{figure}

The most promising measurement for observing the effect
of the large logarithms is the total 
\gstar cross-section, i.e. two-photon scattering of virtual photons
with
 approximately equal virtualities
for the two photons.  
Recent calculations, taking into account higher order effects,
confirm that this remains a gold-plated measurement, which can be 
calculated essentially entirely perturbatively and has a 
sufficiently large 
cross-section. The events are measured by tagging both scattered
electrons. At a 500 GeV \ep collider about 3000 events are expected
per year (200 fb$^{-1}$) and a factor of 3 less in the 
absence of BFKL effects 
in the data~\cite{gammastar}.
 Tagging electrons down to as low angles as possible
(e.g. 25 mrad) is however a crucial requirement for the experiment.
The growth of the cross section as function of 
$W^2$ due to the BFKL effect is shown in Fig.\ref{tdr-phys-qcd:fig_gsgs},
(solid line) and compared with the cross section in absence of BFKL
(dashed line).

Closely related to the \gstar measurement is vector meson production,
e.g \ggx $\rightarrow J/\psi J/\psi$ or (at large $t$) \ggx 
$\rightarrow \rho\rho$,
where the hard scale in the process is given by the $J/\psi$ mass or the 
momentum transfer $t$. $J/\psi$'s can be detected via their decay 
into leptons, and separated from the background through a peak in the 
invariant mass. Approximately 100 fully 
reconstructed 4-muon events are expected for 200 fb$^{-1}$ of
luminosity for a 500 GeV \ep collider~\cite{KWADR}. 
For this channel it is crucial that the decay muons and/or electrons can be 
measured to angles below 10 degrees in the experiment.

A process similar to the 'forward jets' at HERA 
 can be studied at a linear collider in $e\gamma$ 
scattering, with a forward jet produced 
in the direction of the real photon. 
 The measurements can reach out to smaller $x$ 
values than 
presently reachable at HERA, due to the more favourable kinematics
of the final state
~\cite{contreras}.

Finally the processes $e^+e^- \rightarrow e^+e^-\gamma X$ and  
$\gamma\gamma \rightarrow \gamma X$ have 
been studied~\cite{evanson}, and found to be very sensitive to BFKL
dynamics. Event rates for events with photons with energy larger than
5 GeV and $p_T$ larger than 1 GeV are large. At an \ep 
collider several thousand events will be collected per year, 
while at a photon collider the event rate is about a factor ten larger.

In all, the study of these processes will provide new fundamental
insight in small $x$ QCD physics.

\section{Conclusion}
Future linear \ep and \ggx colliders offer a great opportunity 
to study  photon interactions and QCD processes in detail.

\end{document}